\documentclass[11pt]{article}
\usepackage{amsmath}
\usepackage{amsfonts}
\usepackage{amssymb}
\begin{document}

\date{}
\title{\textbf{Higher order relations in Fedosov supermanifolds}}
\author{\textsc{P.M.~Lavrov}\thanks{E-mail: lavrov@tspu.edu.ru} and
\textsc{O.V.~Radchenko}\thanks{E-mail: radchenko@tspu.edu.ru}\\
\\\textit{Tomsk State Pedagogical University,}
\\\textit{634041 Tomsk, Russia}}
\maketitle

\begin{quotation}
\noindent \normalsize Higher order relations existing in normal
coordinates between affine extensions of the curvature tensor and
basic objects for any Fedosov supermanifolds are derived. Representation
of these relations in general coordinates is discussed.
\end{quotation}

\section{Introduction}

Fedosov supermanifolds are a special kind of
supermanifolds introduced by Berezin \cite{Ber} and studied in
details by DeWitt \cite{DeWitt}. They are introduced 
as even or odd symplectic supermanifolds endowed with a symmetric
connection which respects given symplectic structure. In even case
they can be considered as natural extension of Fedosov manifolds
\cite{F,fm} in supersymmetric case. In odd case there is no analog
for them  in differential geometry on manifolds. Note that modern Quantum Field Theory 
involves symplectic supermanifolds to formulate quantization procedures.
The well-known quantization
method proposed by Batalin and Vilkovisky \cite{bv} is based on
geometry of  odd symplectic supermanifolds \cite{Wit}. 
In turn the deformation quantization \cite{F} can be formulated for any even symplectic
supermanifolds (see \cite{B,BHW}).
Simple kind of Fedosov supermanifolds  has been already appeared in physical literature.
Namely, flat even Fedosov supermanifolds have been used to
construct coordinate free quantization procedure \cite{bt} and
triplectic quantization method \cite{BM,mod3pl} in general
coordinates \cite{gl}. 

Systematic investigation of basic properties of even and odd
Fedosov supermanifolds has been started in \cite{gl2} and
continued in \cite{gl1}. In particular, some basic difference in
even and odd Fedosov supermanifolds has been found which can be
expressed in terms of the scalar curvature $K$. Namely, for any
even Fedosov supermanifold the scalar curvature, as in usual
differential geometry, is equal to zero while for odd Fedosov
supermanifolds it is, in general, non-trivial. Moreover there
exist the relations between a supersymplectic structure, a
connection (Christoffel symbols) and the curvature tensor found in 
\cite{gl1} in the lowest (first and second) orders which are defined by orders of affine
extension of connection on supermanifolds.

The goal of the present paper is to study higher order relations
existing among the Christoffel symbols, symplectic structure and
the curvature tensor and to find a fundamental origin of all these
relations.

The paper is organized as follows. In Sect.~2, we  give the notion
of even (odd) Fedosov supermanifolds and of even (odd) symplectic
curvature tensor. In Sect.~3, we consider affine extensions of
the Christoffel symbols and tensors on a supermanifold. In
Sect.~4, we consider relations existing among  affine extensions
of the Christoffel symbols, symplectic structure and the curvarute
tensor of the first and second orders. In Sect.~5, we study
relations of the third order for objects listed in Sect.~4. In
Sect.~6, we present relations obtained in normal coordinates using
general coordinates on  a supermanifold. In Sect.~7 we give a few
concluding remarks.

We use the condensed notation suggested by DeWitt \cite{condnot}.
Derivatives with respect to the coordinates $x^i$ are understood
as acting from the right and for them the notation $
A_{,i}={\partial_r A}/{\partial x^i}$ is used.  Covariant
derivatives are understood as acting to the right with the
notation $A_{;i}=A\nabla_i$. The Grassmann parity of any quantity
$A$ is denoted by $\epsilon (A)$.
\\

\section{Fedosov supermanifolds}

Consider an even (odd) symplectic supermanifold,
$(M,\omega)$ with an even (odd) symplectic structure $\omega,\;
\epsilon(\omega) = 0$ (or $1$). Let us equip $(M,\omega)$ with  a covariant derivative
(connection) $\nabla$ (or $\Gamma$) which preserves the symplectic structure $\omega$,
$\omega\nabla=0$.  In a coordinate basis this requirement reads
\begin{eqnarray}
\label{covomiv} \omega_{ij,k}-\omega_{im}\Gamma^m_{\;\;jk}+
\omega_{jm}\Gamma^m_{\;\;ik}(-1)^{\epsilon_i\epsilon_j}=0,\quad
\omega_{ij}=-\omega_{ji}(-1)^{\epsilon_i\epsilon_j}.
\end{eqnarray}
If, in addition, $\Gamma$ is symmetric
$\Gamma^i_{\;\;jk}=\Gamma^i_{\;\;kj}(-1)^{\epsilon_k\epsilon_j}$
then  the triple $(M,\omega,\Gamma)$ is defined as a Fedosov
supermanifold.

The curvature tensor of a symplectic
connection with all indices lowered,
\begin{eqnarray}
\label{Rs}
\nonumber
R_{imjk}&=&\omega_{in}R^n_{\;\;mjk}
 =-\omega_{in}\Gamma^n_{\;\;mj,k}+
\omega_{in}\Gamma^n_{\;\;mk,j}(-1)^{\epsilon_j\epsilon_k}+\\
&&
\Gamma_{ijn}\Gamma^n_{\;\;mk}(-1)^{\epsilon_j\epsilon_m}-
\Gamma_{ikn}\Gamma^n_{\;\;mj}
(-1)^{\epsilon_k(\epsilon_m+\epsilon_j)}\,,
\end{eqnarray}
obeys the following symmetry properties \cite{gl2}
\begin{eqnarray}
\label{Rans} R_{ijkl}=-(-1)^{\epsilon_k\epsilon_l}R_{ijlk},\quad
R_{ijkl}=(-1)^{\epsilon_i\epsilon_j}R_{jikl}.
\end{eqnarray}
In (\ref{Rs})  we used the notation
\begin{eqnarray}
\nonumber
\label{G} \Gamma_{ijk}=\omega_{in}\Gamma^n_{\;\;jk},\quad
\epsilon(\Gamma_{ijk})=\epsilon(\omega)+
\epsilon_i+\epsilon_j+\epsilon_k\,.
\end{eqnarray}
Using definition of tensor field $\omega^{ij}$ inverse to the symplectic
structure $\omega_{ij}$
\begin{eqnarray}
\nonumber
\label{winv}
\omega_{in}\omega^{nj}(-1)^{\epsilon_i+\epsilon(\omega)(\epsilon_i+\epsilon_n)}=\delta^j_i,\quad
\omega^{ij}=-\omega^{ji}(-1)^{\epsilon_i\epsilon_j+\epsilon(\omega)},
\end{eqnarray}
one obtains
\begin{eqnarray}
\label{wG}
\omega_{in}\Gamma^n_{\;\;jk,l}=\Gamma_{ijk,l}-
\Gamma_{inl}\Gamma^n_{\;\;jk}
(-1)^{\epsilon_l(\epsilon_n+\epsilon_j+\epsilon_k)}+
\Gamma_{nil}\Gamma^n_{\;\;jk}
(-1)^{\epsilon_l(\epsilon_n+\epsilon_j+\epsilon_k)+\epsilon_n\epsilon_i}
\end{eqnarray}
and, therefore, the following representation for the curvature tensor
\begin{eqnarray}
\label{Rl}
R_{ijkl}=-\Gamma_{ijk,l}+\Gamma_{ijl,k}(-1)^{\epsilon_l\epsilon_k}+
\Gamma_{nik}\Gamma^n_{\;\;jl}
(-1)^{\epsilon_k(\epsilon_n+\epsilon_j)+\epsilon_n\epsilon_i}-
\Gamma_{nil}\Gamma^n_{\;\;jk}
(-1)^{\epsilon_l(\epsilon_n+\epsilon_j+\epsilon_k)+\epsilon_n\epsilon_i}.
\end{eqnarray}
The (super) Jacobi identity for $R_{ijkl}$ holds
\begin{eqnarray}
\label{Rjac1} R_{ijkl}(-1)^{\epsilon_j\epsilon_l}
+R_{iljk}(-1)^{\epsilon_l\epsilon_k}
+R_{iklj}(-1)^{\epsilon_k\epsilon_j}=0\,.
\end{eqnarray}
For any even (odd) symplectic connection there holds the identity
\cite{gl2}
\begin{eqnarray}
\label{Rjac2} R_{ijkl}
+R_{lijk}(-1)^{\epsilon_l(\epsilon_k+\epsilon_j+\epsilon_i)}
+R_{klij}(-1)^{(\epsilon_k+\epsilon_l)(\epsilon_i+\epsilon_j)}
+
R_{jkli}(-1)^{\epsilon_i(\epsilon_j+\epsilon_k+\epsilon_l)}=0.
\end{eqnarray}
In the identity (\ref{Rjac2}) the components of the symplectic
curvature tensor occur with cyclic permutations of all the indices
(on $R$). However, the pre-factors depending on the Grassmann
parities of indices are not obtained by cyclic permutation as in
case of the Jacobi identity (\ref{Rjac1}) but by permutation of
indices from given set to initial one.

\section{Affine extensions of tensors on supermanifolds}

In Ref. \cite{fm} the virtues of using normal coordinates for
studying the properties of Fedosov manifolds was demonstrated.
Normal coordinates
$\{y^i\}$ within a point $p\in M$ can be introduced by using the  geodesic
equations as those local coordinates which satisfy the relations
($p$ corresponds to $y=0$)
\begin{eqnarray}
\label{Chsy}
 \Gamma^i_{\;jk}(y)\,y^k\,y^j = 0,
 \quad\epsilon(\Gamma_{ijk})=\epsilon(\omega)+\epsilon_i
+\epsilon_j+\epsilon_k.
\end{eqnarray}
It follows from (\ref{Chsy}) and the symmetry properties of $\Gamma_{ijk}$
w.r.t.~$(j\,k)$ that
\begin{eqnarray}
\nonumber
\label{Chsy0}
\Gamma_{ijk}(0)= 0.
\end{eqnarray}
In normal coordinates there exist additional relations at $p$
containing the partial derivatives of $\Gamma_{ijk}$. Namely,
consider the Taylor expansion of $\Gamma_{ijk}(y)$ at $y=0$,
\begin{eqnarray}
\label{TCh}
\Gamma_{ijk}(y)=\sum_{n=1}^{\infty}\frac{1}{n!}
A_{ijkj_1...j_n}y^{j_n}\cdot\cdot\cdot y^{j_1},\quad
{\rm where} \quad
A_{ijkj_1\ldots j_n}=A_{ijkj_1\ldots j_n}(p)=\left.
\frac{\partial_r^n \Gamma_{ijk}}
{\partial y^{j_1}\ldots \partial y^{j_n}}\right|_{y=0}
\end{eqnarray}
is called an affine extension of $\Gamma_{ijk}$ of order
$n=1,2,\ldots$ .
The symmetry properties of $A_{ijkj_1\ldots j_n}$ are evident from
their definition (\ref{TCh}), namely, they are (generalized)
symmetric w.r.t.~$(j\,k)$ as well as $(j_1\ldots j_n)$. The set of
all affine extensions of $\Gamma_{ijk}$ uniquely defines a
symmetric connection according to (\ref{TCh}) and satisfy an
infinite sequence of identities \cite{gl2}. In the lowest
nontrivial orders they have the form
\begin{eqnarray}
\label{r3} A_{ijkl}+A_{ijlk}(-1)^{\epsilon_k\epsilon_l} +
A_{iklj}(-1)^{\epsilon_j(\epsilon_l+\epsilon_k)}=0,
\end{eqnarray}
\begin{eqnarray}
\label{r6} \nonumber &&A_{ijklm}+
A_{ijlkm}(-1)^{\epsilon_k\epsilon_l} +
A_{ikljm}(-1)^{\epsilon_j(\epsilon_l+\epsilon_k)}\\
&&+A_{ijmkl}(-1)^{\epsilon_m(\epsilon_l+\epsilon_k)}+
A_{ilmjk}(-1)^{(\epsilon_j+\epsilon_k)(\epsilon_m+\epsilon_l)} +
A_{ikmjl}(-1)^{\epsilon_j(\epsilon_m+\epsilon_k)+\epsilon_m\epsilon_l}=0
\end{eqnarray}
and
\begin{eqnarray}
\label{r10} \nonumber &&A_{ijklmn}+
A_{ijlkmn}(-1)^{\epsilon_k\epsilon_l} +
A_{ikljmn}(-1)^{\epsilon_j(\epsilon_l+\epsilon_k)}\\
\nonumber
 &&+A_{ijmkln}(-1)^{\epsilon_m(\epsilon_l+\epsilon_k)}+
A_{ilmjkn}(-1)^{(\epsilon_j+\epsilon_k)(\epsilon_m+\epsilon_l)}
+A_{ikmjln}(-1)^{\epsilon_j(\epsilon_m+\epsilon_k)+\epsilon_m\epsilon_l}\\
\nonumber
&&+A_{ijnklm}(-1)^{\epsilon_n(\epsilon_l+\epsilon_k+\epsilon_m)}+
A_{iknjlm}(-1)^{\epsilon_j(\epsilon_n+\epsilon_k)+\epsilon_n(\epsilon_l+\epsilon_m)}\\
&&+
A_{ilnjkm}(-1)^{(\epsilon_j+\epsilon_k)(\epsilon_n+\epsilon_l)+\epsilon_m\epsilon_n}+
A_{imnjkl}(-1)^{(\epsilon_j+\epsilon_k+\epsilon_l)(\epsilon_m+\epsilon_n)}
 =0.
\end{eqnarray}

 Analogously, the affine extensions of an arbitrary tensor
$T=(T^{i_1...i_k}_{\;\;\;\;\;\;\;\;\;\;m_1...m_l})$ on $M$ are
defined as  tensors on $M$ whose components at $p\in M$ in the
local coordinates $(x^1,\ldots,x^{2N})$ are given by the formula
\begin{eqnarray}
\nonumber
\label{AfT}
T^{i_1...i_k}_{\;\;\;\;\;\;\;\;m_1...m_l,j_1...j_n} \equiv
T^{i_1...i_k}_{\;\;\;\;\;\;\;\;m_1...m_l,j_1...j_n}(0)=\left.
\frac{\partial_r^n
T^{i_1...i_k}_{\;\;\;\;\;\;\;\;\;m_1...m_l}}
{\partial y^{j_1}...\partial y^{j_n}}\right|_{y=0}
\end{eqnarray}
where $(y^1,\ldots,y^{2N})$ are normal coordinates associated with
$(x^1,\ldots,x^{2N})$ at $p$. The first extension of any tensor
coincides with its covariant derivative because
$\Gamma^i_{\;\;jk}(0)=0$ in normal coordinates.

In the following, any relation containing affine extensions are to
be understood as holding in a neighborhood $U$ of an arbitrary
point $p \in M$. Let us also observe the convention that, if a
relation holds for arbitrary local coordinates, the arguments of
the related quantities will be suppressed. The order of relations
is defined by the order of affine extension of  $\Gamma_{ijk}$ 
entering in the relations.

\section{First and second order relations}

For a given Fedosov supermanifold $(M,\omega, \Gamma)$, 
symmetric connection $\Gamma$ respects the symplectic structure
$\omega$ \cite{gl}:
\begin{eqnarray}
\label{sc}
\omega_{ij,k}=\Gamma_{ijk}-\Gamma_{jik}(-1)^{\epsilon_i\epsilon_j}.
\end{eqnarray}
Therefore, among the affine extensions of
$\omega_{ij}$ and $\Gamma_{ijk}$ there must exist some relations.
Introducing the affine extensions of $\omega_{ij}$
in the normal coordinates $(y^1,\ldots,y^{2N})$ at $p\in M$
according to
\begin{eqnarray}
\nonumber
\label{Tew} \omega_{ij}(y) =\sum_{n=0}^{\infty}\frac{1}{n!}
\omega_{ij,j_1...j_n}(0)\, y^{j_n}\cdots y^{j_1},
\end{eqnarray}
using the symmetry properties of $\omega_{ij,j_1 \ldots j_n}(0)$
and the fact $\omega_{ij,k}(0)=0$ one easily obtains the Taylor
expansion for $\omega_{ij,k}$:
\begin{eqnarray}
\label{Tedw} \omega_{ij,k}(y) =\sum_{n=1}^{\infty}\frac{1}{n!}
\omega_{ij,kj_1...j_n}(0)\, y^{j_n}\cdots y^{j_1}.
\end{eqnarray}
Taking into account (\ref{sc}) and comparing
(\ref{TCh}) and (\ref{Tedw}) we obtain
\begin{eqnarray}
\label{wA} \omega_{ij,kj_1\ldots j_n}(0)=A_{ijkj_1\ldots j_n}-
A_{jikj_1\ldots j_n}(-1)^{\epsilon_i\epsilon_j};
\end{eqnarray}
in particular,
\begin{eqnarray}
\label{wA1} \omega_{ij,kl}(0)=A_{ijkl}-
A_{jikl}(-1)^{\epsilon_i\epsilon_j}.
\end{eqnarray}

Now, consider the curvature tensor $R_{ijkl}$ in the normal
coordinates at $p\in M$. Then, due to $\Gamma_{ijk}(p)=0$, we
obtain the following representation of the curvature tensor in
terms of the affine extensions of $\Gamma_{ijk}$
\begin{eqnarray}
\label{Rnc}
R_{ijkl}(0)= -
A_{ijkl}+A_{ijlk}(-1)^{\epsilon_k\epsilon_l}.
\end{eqnarray}
From (\ref{r3}) 
and (\ref{Rnc})
a relation containing the curvature
tensor and the first affine extension of $\Gamma$ can be derived.
Indeed, 
the desired relation obtains as follows
\begin{eqnarray}
\label{RA1}
A_{ijkl}\equiv \Gamma_{ijk,l}(0) =
-\frac{1}{3}\left[R_{ijkl}(0)+R_{ikjl}(0)(-1)^{\epsilon_k\epsilon_j}\right],
\end{eqnarray}
where the antisymmetry (\ref{Rans}) of the curvature tensor were
used.

Taking into account  (\ref{wA1}) and (\ref{Rnc}) it follows the relation between
the second order affine extension of symplectic structure and the
symplectic curvature tensor.  Indeed, using the Jacobi identity
(\ref{Rjac1}), we obtain
\begin{eqnarray}
\label{wAR1}
\omega_{ij,kl}(0)=A_{ijkl}-
A_{jikl}(-1)^{\epsilon_i\epsilon_j} =\frac{1}{3}R_{klij}(0)
(-1)^{(\epsilon_i+\epsilon_j)(\epsilon_k+\epsilon_l)},
\end{eqnarray}
Symmetry properties of $R_{klij}(0)$ and $\omega_{ij,kl}(0)$ are
in accordance with this relation.

Having representation of the first order affine extension of the
Christoffel symbols, $A_{ijkl}$, in terms of the curvature tensor,
$R_{ijkl}$, the relation (\ref{r3}) can be considered as
consequence of the Jacobi identity (\ref{Rjac1}) or of the
antisymmetry property of $R_{ijkl}$ with respect to two last
indices (\ref{Rans}).

Differentiating both sides of (\ref{Rl}) w.r.t.~$y^m$, taking the limit
$y\rightarrow 0$ and observing that, because of
$\Gamma_{ijk}(0)=0$, the first extension of the symplectic
structure vanishes,$\omega_{ij,k}(0)=0$, we have
\begin{eqnarray}
\label{RA2} R_{ijkl,m}(0)=-A_{ijklm} +
A_{ijlkm}(-1)^{\epsilon_l\epsilon_k}.
\end{eqnarray}
That relation will be used to eliminate within the relation
(\ref{r6}) all the extensions of the Christoffel symbols in favor
of $A_{ijklm}$
and to obtain the following representation of the second order
affine extension of $\Gamma_{ijk}$ in terms of first order
derivatives of the curvature tensor
\begin{eqnarray}
\label{A2R} \nonumber A_{ijklm} &=& -\frac{1}{6}\Big[\,
2R_{ijkl,m}(0) + R_{ijkm,l}(0)(-1)^{\epsilon_m\epsilon_l}+
R_{ikjl,m}(0)(-1)^{\epsilon_j\epsilon_k}\\
&& \qquad +
R_{ikjm,l}(0)(-1)^{\epsilon_j\epsilon_k+\epsilon_m\epsilon_l}+
R_{iljm,k}(0)(-1)^{\epsilon_j\epsilon_l+\epsilon_k(\epsilon_m+\epsilon_l)}
\Big].
\end{eqnarray}
It is easy to check that the representation for $A_{ijklm}$
(\ref{A2R}) is accordingly with the relation (\ref{r6}) due to the
antisymmetry property of the curvature tensor $R_{ijkl}$.

In turn from (\ref{wA}) we have
\begin{eqnarray}
\label{wA3}
\nonumber
\omega_{ik,jlm}(0)=A_{ikjlm}-A_{kijlm}(-1)^{\epsilon_i\epsilon_k},
\end{eqnarray}
and therefore we get
\begin{align}
\label{wA3R}
&\omega_{ij,klm}(0)=-\frac{1}{6}\Big[
 R_{ikjl,m}(0)(-1)^{\epsilon_j\epsilon_k}
+R_{ikjm,l}(0)(-1)^{\epsilon_j\epsilon_k+\epsilon_m\epsilon_l}
+R_{iljm,k}(0)(-1)^{\epsilon_j\epsilon_l+\epsilon_k(\epsilon_l+\epsilon_m)}
\\ \nonumber 
& \qquad\quad
-R_{jkil,m}(0)(-1)^{\epsilon_i(\epsilon_k+\epsilon_j)}
-R_{jkim,l}(0)(-1)^{\epsilon_m\epsilon_l+\epsilon_i(\epsilon_j+\epsilon_k)}
-R_{jlim,k}(0)(-1)^{\epsilon_k(\epsilon_m+\epsilon_l)
+\epsilon_i(\epsilon_j+\epsilon_l)}\Big]
\end{align}
as the representation of the third order affine extensions of
$\omega_{ij}$ in terms of the first order affine extension of the
symplectic curvature tensor.

Having the representation (\ref{wA3R})
 we can consider the consequences which follow from the symmetry
properties of $\omega_{ik,jlm}$,
\begin{eqnarray}
\nonumber
\label{ws3}
\omega_{ik,jlm}=\omega_{ik,ljm}(-1)^{\epsilon_l\epsilon_j}
\end{eqnarray}
and  obtain the following identities for the first affine
extension of the curvature tensor
\begin{align}
\label{IIRy} \nonumber
&R_{mjik,l}(0)(-1)^{\epsilon_j(\epsilon_i+\epsilon_k)}-
R_{mijl,k}(0)(-1)^{\epsilon_k(\epsilon_l+\epsilon_j)}\\
&+R_{mkjl,i}(0)(-1)^{\epsilon_i(\epsilon_j+\epsilon_k+\epsilon_l)}-
R_{mlik,j}(0)(-1)^{\epsilon_l(\epsilon_i+\epsilon_j+\epsilon_k)}=0.
\end{align}
Note that the identity (\ref{IIRy}) cannot be considered as
relation containing a cyclic permutation of four indices of the
first order extension of the curvature tensor.

\section{Third  order relations}

Beginning with the third order relations we meet a new feature
concerning representation of affine extensions of the curvature
tensor and the symplectic structure in terms of affine extensions
of the Christoffel symbols. This feature is connected with
nonlinear  dependence in contrast with relations of the first and
second orders. Indeed, from (\ref{Rl}) it can be derived the
following representation for the second order extension of the
curvature tensor
\begin{eqnarray}
\label{Re2}
R_{ijkl,mn}(0)=-A_{ijklmn}+A_{ijlkmn}(-1)^{\epsilon_k\epsilon_l}+N_{ijklmn},
\end{eqnarray}
where
\begin{eqnarray}
\label{N}
N_{ijklmn}=T_{ijklmn}+T_{ijklnm}(-1)^{\epsilon_n\epsilon_m}-
T_{ijlkmn}(-1)^{\epsilon_k\epsilon_l}-T_{ijlknm}(-1)^{\epsilon_k\epsilon_l+\epsilon_n\epsilon_m},
\end{eqnarray}
\begin{eqnarray}
\label{T}
T_{ijklmn}=A_{sikm}A^s_{\;\;jln}(-1)^{\epsilon_s\epsilon_i+\epsilon_k(\epsilon_s+\epsilon_j)+
\epsilon_m(\epsilon_l+\epsilon_j+\epsilon_s)}
\end{eqnarray}
is quadratic in the first order extension of the Christoffel
symbols. In (\ref{N}) we used the notation
\begin{eqnarray}
\label{Aup}
A^i_{\;\;jkl}=\omega^{ip}A_{pjkl}(-1)^{\epsilon_p+\epsilon(\omega)(\epsilon_p+\epsilon_i)}=
-\frac{1}{3}[R^i_{\;\;jkl}(0)+R^i_{\;\;kjl}(0)(-1)^{\epsilon_k\epsilon_j}].
\end{eqnarray}
From (\ref{T}) it follows the following symmetry properties
\begin{eqnarray}
\label{Ts1}
T_{ijklmn}&=&T_{ilkjmn}(-1)^{\epsilon_l(\epsilon_k+\epsilon_j)+\epsilon_j\epsilon_k},\\
\label{Ts2}
T_{ijklmn}&=&T_{kjilmn}(-1)^{\epsilon_k(\epsilon_i+\epsilon_j)+\epsilon_i\epsilon_j},\\
\label{Ts3}
T_{ijklmn}&=&-T_{jilknm}(-1)^{\epsilon_i\epsilon_j+\epsilon_k\epsilon_l+\epsilon_m\epsilon_n}.
\end{eqnarray}

Taking into account the symmetry properties of $A_{ijklmn}$ it
follows from (\ref{Re2})
\begin{eqnarray}
\label{Ae3} \nonumber A_{ijklmn}(-1)^{\epsilon_k\epsilon_l}&=&
R_{ijkl,mn}(0) + A_{ijklmn}-N_{ijklmn}, \\
\nonumber A_{ijmkln}(-1)^{\epsilon_k\epsilon_m}&=&
R_{ijkm,ln}(0) + A_{ijkmln}-N_{ijkmln}= \\
\nonumber && R_{ijkm,ln}(0) +
A_{ijklmn}(-1)^{\epsilon_m\epsilon_l}-N_{ijkmln},\\
\nonumber A_{ijnklm}(-1)^{\epsilon_k\epsilon_n}&=&
R_{ijkn,lm}(0) + A_{ijknlm}-N_{ijknlm}= \\
\nonumber && R_{ijkn,lm}(0) +
A_{ijklmn}(-1)^{\epsilon_l\epsilon_n+\epsilon_m\epsilon_n}-N_{ijknlm},\\
\nonumber  A_{ikljmn}(-1)^{\epsilon_l\epsilon_j}&=&
R_{ikjl,mn}(0) + A_{ikjlmn}-N_{ikjlmn}= \\
\nonumber && R_{ikjl,mn}(0) +
A_{ijklmn}(-1)^{\epsilon_j\epsilon_k}-N_{ikjlmn},
\end{eqnarray}
\begin{eqnarray}
\nonumber A_{ikmjln}(-1)^{\epsilon_m\epsilon_j}&=&
R_{ikjm,ln}(0) + A_{ikjmln}-N_{ikjmln}= \\
\nonumber && R_{ikjm,ln}(0) +
A_{ijklmn}(-1)^{\epsilon_k\epsilon_j+\epsilon_m\epsilon_l}-N_{ikjmln},\\
\nonumber A_{iknjlm}(-1)^{\epsilon_j\epsilon_n}&=&
R_{ikjn,lm}(0) + A_{ikjnlm}-N_{ikjnlm}= \\
\nonumber && R_{ikjn,lm}(0) +
A_{ijklmn}(-1)^{\epsilon_j\epsilon_k+\epsilon_n(\epsilon_m+\epsilon_l)}-N_{ikjnlm},\\
\nonumber A_{ilmjkn}(-1)^{\epsilon_j\epsilon_m}&=&
R_{iljm,kn}(0) + A_{iljmkn}-N_{iljmkn}= \\
\nonumber && R_{iljm,kn}(0) +
A_{ijlkmn}(-1)^{\epsilon_j\epsilon_l+\epsilon_k\epsilon_m}-N_{iljmkn}=\\
\nonumber &&R_{iljm,kn}(0)+(R_{ijkl,mn}+A_{ijklmn}-N_{ijklmn})
(-1)^{\epsilon_j\epsilon_l+\epsilon_k\epsilon_l+\epsilon_k\epsilon_m}
-N_{iljmkn},\\
\nonumber A_{ilnjkm}(-1)^{\epsilon_j\epsilon_n}&=&
R_{iljn,km}(0) + A_{iljnkm}-N_{iljnkm}= \\
\nonumber && R_{iljn,km}(0) +
A_{ijlkmn}(-1)^{\epsilon_j\epsilon_l+\epsilon_k\epsilon_n+\epsilon_m\epsilon_n}-N_{iljnkm}=\\
\nonumber &&R_{iljn,km}(0)+(R_{ijkl,mn}+A_{ijklmn}-N_{ijklmn})
(-1)^{\epsilon_j\epsilon_l+\epsilon_k\epsilon_l+\epsilon_k\epsilon_n+\epsilon_m\epsilon_n}
-N_{iljnkm},\\
\nonumber
 A_{imnjkl}(-1)^{\epsilon_j\epsilon_n}&=&
R_{imjn,kl}(0) + A_{imjnkl}-N_{imjnkl}= \\
\nonumber && R_{imjn,kl}(0) +
A_{ijmknl}(-1)^{\epsilon_j\epsilon_m+\epsilon_k\epsilon_n}-N_{imjnkl}=\\
\nonumber &&R_{iljn,km}(0)+(R_{ijkl,mn}+A_{ijklmn}-N_{ijklmn})
(-1)^{\epsilon_j\epsilon_m+\epsilon_k\epsilon_m+\epsilon_k\epsilon_n}
-N_{imjnkl}.
\end{eqnarray}
Putting them into the identity (\ref{r10}) we obtain
\begin{eqnarray}
\label{Ae3}
\nonumber
&&10A_{ijklmn}+3 R_{ijkl,mn} + 2
R_{ijkm,ln}(-1)^{\epsilon_m\epsilon_l}+R_{ikjl,mn}(-1)^{\epsilon_j\epsilon_k}+\\
\nonumber
&&R_{ijkn,lm}(-1)^{\epsilon_n(\epsilon_m+\epsilon_l)}+
R_{ikjn,lm}(-1)^{\epsilon_j\epsilon_k+\epsilon_n(\epsilon_m+\epsilon_l)}
+R_{ikjm,ln}(-1)^{\epsilon_j\epsilon_k+\epsilon_m\epsilon_l}+\\
\nonumber
&&R_{iljm,kn}(-1)^{\epsilon_j\epsilon_l+\epsilon_k\epsilon_l+\epsilon_m\epsilon_k}
+R_{iljn,km}(-1)^{\epsilon_l(\epsilon_k+\epsilon_j)+\epsilon_n(\epsilon_k+\epsilon_m)}+
R_{imjn,kl}(-1)^{\epsilon_j\epsilon_m+(\epsilon_l+\epsilon_k)(\epsilon_n+\epsilon_m)}-\\
\nonumber
&&3N_{ijklmn}- N_{ijkmln}(-1)^{\epsilon_m\epsilon_l}-
N_{ijknlm}(-1)^{\epsilon_n(\epsilon_m+\epsilon_l)}-
N_{ikjlmn}(-1)^{\epsilon_j\epsilon_k}-\\
\nonumber
&&N_{ikjmln}(-1)^{\epsilon_j\epsilon_k+\epsilon_m\epsilon_l}-
N_{ikjnlm}(-1)^{\epsilon_j\epsilon_k+\epsilon_n(\epsilon_m+\epsilon_l)}-
N_{iljmkn}(-1)^{\epsilon_j\epsilon_l+\epsilon_k\epsilon_l+\epsilon_m\epsilon_k}-\\
\nonumber
&&N_{iljnkm}(-1)^{\epsilon_l(\epsilon_k+\epsilon_j)+\epsilon_n(\epsilon_k+\epsilon_m)}-
N_{ijkmnl}(-1)^{\epsilon_l(\epsilon_m+\epsilon_n)}-
N_{imjnkl}(-1)^{\epsilon_j\epsilon_m+(\epsilon_l+\epsilon_k)(\epsilon_n+\epsilon_m)}=0.
\end{eqnarray}
Therefore, we have the nonlinear representation of the third order extension of the Christoffel symbols
in terms of the curvature tensor
\begin{eqnarray}
\label{Ae3l}
\nonumber
A_{ijklmn}&=&-\frac{1}{10}[3 R_{ijkl,mn} + 2
R_{ijkm,ln}(-1)^{\epsilon_m\epsilon_l}+R_{ikjl,mn}(-1)^{\epsilon_j\epsilon_k}+
R_{ijkn,lm}(-1)^{\epsilon_n(\epsilon_m+\epsilon_l)}+\\
\nonumber
&&R_{ikjn,lm}(-1)^{\epsilon_j\epsilon_k+\epsilon_n(\epsilon_m+\epsilon_l)}
+R_{ikjm,ln}(-1)^{\epsilon_j\epsilon_k+\epsilon_m\epsilon_l}+
R_{iljm,kn}(-1)^{\epsilon_j\epsilon_l+\epsilon_k\epsilon_l+\epsilon_m\epsilon_k}
+\\
\nonumber
&&R_{iljn,km}(-1)^{\epsilon_l(\epsilon_k+\epsilon_j)+\epsilon_n(\epsilon_k+\epsilon_m)}+
R_{imjn,kl}(-1)^{\epsilon_j\epsilon_m+(\epsilon_l+\epsilon_k)(\epsilon_n+\epsilon_m)}-\\
\nonumber
&&3N_{ijklmn}- N_{ijkmln}(-1)^{\epsilon_m\epsilon_l}-
N_{ijknlm}(-1)^{\epsilon_n(\epsilon_m+\epsilon_l)}-
N_{ikjlmn}(-1)^{\epsilon_j\epsilon_k}-\\
\nonumber
&&N_{ikjmln}(-1)^{\epsilon_j\epsilon_k+\epsilon_m\epsilon_l}-
N_{ikjnlm}(-1)^{\epsilon_j\epsilon_k+\epsilon_n(\epsilon_m+\epsilon_l)}-
N_{iljmkn}(-1)^{\epsilon_j\epsilon_l+\epsilon_k\epsilon_l+\epsilon_m\epsilon_k}-\\
\nonumber
&&N_{iljnkm}(-1)^{\epsilon_l(\epsilon_k+\epsilon_j)+\epsilon_n(\epsilon_k+\epsilon_m)}-
N_{ijkmnl}(-1)^{\epsilon_l(\epsilon_m+\epsilon_n)}-\\
&&N_{imjnkl}(-1)^{\epsilon_j\epsilon_m+(\epsilon_l+\epsilon_k)(\epsilon_n+\epsilon_m)}].
\end{eqnarray}
Taking into account the relation between affine extensions of the symplectic
structure and the Christoffel symbols
\begin{eqnarray}
\label{w4}
\nonumber
\omega_{ij,klmn}=A_{ijklmn}-A_{jiklmn}(-1)^{\epsilon_i\epsilon_j},
\end{eqnarray}
we derive the following formula for the forth order affine extension of the
symplectic structure in terms of the curvature tensor

\begin{eqnarray}
\label{wR2}
\nonumber
\omega_{ij,klmn}&=&
-\frac{1}{10}\big[R_{ikjl,mn}(-1)^{\epsilon_j\epsilon_k}-R_{jkil,mn}(-1)^{\epsilon_i(\epsilon_k+\epsilon_j)}+\\
\nonumber
&&R_{imjn,kl}(-1)^{\epsilon_j\epsilon_m+(\epsilon_l+\epsilon_k)(\epsilon_n+\epsilon_m)}-
R_{jmin,kl}(-1)^{\epsilon_i(\epsilon_m+\epsilon_j)+(\epsilon_l+\epsilon_k)(\epsilon_n+\epsilon_m)}+
\\
\nonumber
&&R_{ikjn,lm}(-1)^{\epsilon_j\epsilon_k+\epsilon_n(\epsilon_m+\epsilon_l)}-
R_{jkin,lm}(-1)^{\epsilon_i(\epsilon_k+\epsilon_j)+\epsilon_n(\epsilon_m+\epsilon_l)}
+\\
\nonumber
&&R_{ikjm,ln}(-1)^{\epsilon_j\epsilon_k+\epsilon_m\epsilon_l}-
R_{jkim,ln}(-1)^{\epsilon_i(\epsilon_k+\epsilon_j)+\epsilon_m\epsilon_l}+\\
\nonumber
&&
R_{iljm,kn}(-1)^{\epsilon_j\epsilon_l+\epsilon_k(\epsilon_l+\epsilon_m)}-
R_{jlim,kn}(-1)^{\epsilon_i(\epsilon_l+\epsilon_j)+\epsilon_k(\epsilon_l+\epsilon_m)}
+\\
\nonumber
&&R_{iljn,km}(-1)^{\epsilon_l(\epsilon_k+\epsilon_j)+\epsilon_n(\epsilon_k+\epsilon_m)}-
R_{jlin,km}(-1)^{\epsilon_i\epsilon_j+\epsilon_l(\epsilon_k+\epsilon_i)+\epsilon_n(\epsilon_k+\epsilon_m)}+\\
\nonumber
&&R_{imjn,kl}(-1)^{\epsilon_j\epsilon_m+(\epsilon_l+\epsilon_k)(\epsilon_n+\epsilon_m)}-
R_{jmin,kl}(-1)^{\epsilon_i(\epsilon_m+\epsilon_j)+(\epsilon_l+\epsilon_k)(\epsilon_n+\epsilon_m)}-\\
\nonumber
&&
N_{ikjlmn}(-1)^{\epsilon_j\epsilon_k}+N_{jkilmn}(-1)^{\epsilon_i(\epsilon_k+\epsilon_j)}-\\
\nonumber
&&N_{ikjmln}(-1)^{\epsilon_j\epsilon_k+\epsilon_m\epsilon_l}+
N_{jkimln}(-1)^{\epsilon_i(\epsilon_k+\epsilon_j)+\epsilon_m\epsilon_l}-\\
\nonumber
&&N_{ikjnlm}(-1)^{\epsilon_j\epsilon_k+\epsilon_n(\epsilon_m+\epsilon_l)}+
N_{jkinlm}(-1)^{\epsilon_i(\epsilon_k+\epsilon_j)+\epsilon_n(\epsilon_m+\epsilon_l)}-\\
\nonumber
&&
N_{iljmkn}(-1)^{\epsilon_j\epsilon_l+\epsilon_k(\epsilon_l+\epsilon_m)}+
N_{jlimkn}(-1)^{\epsilon_i(\epsilon_l+\epsilon_j)+\epsilon_k(\epsilon_l+\epsilon_m)}-\\
\nonumber
&&N_{iljnkm}(-1)^{\epsilon_l(\epsilon_k+\epsilon_j)+\epsilon_n(\epsilon_k+\epsilon_m)}+
N_{jlinkm}(-1)^{\epsilon_i\epsilon_j+\epsilon_l(\epsilon_k+\epsilon_i)+\epsilon_n(\epsilon_k+\epsilon_m)}-
\\
&&N_{imjnkl}(-1)^{\epsilon_j\epsilon_m+(\epsilon_l+\epsilon_k)(\epsilon_n+\epsilon_m)}+
N_{jminkl}(-1)^{\epsilon_i(\epsilon_j+\epsilon_m)+(\epsilon_l+\epsilon_k)(\epsilon_n+\epsilon_m)}\big].
\end{eqnarray}

We have already found that symmetry properties of the third order
affine extension of symplectic structure expressed in terms of the
curvature tensor (\ref{wA3}) led to the new identity (\ref{IIRy})
for the curvature tensor. A natural question appears:  Are there
some new identities containing the second order affine extension
of the curvature tensor as consequences of the representation
(\ref{wR2}) and symmetry properties
\begin{eqnarray}
\label{w4s}
\nonumber
\omega_{ij,klmn}-\omega_{ij,lkmn}(-1)^{\epsilon_k\epsilon_l}=0
\end{eqnarray}
of the fourth order affine extension of the symplectic structure?
We shall prove that the answer is negative. Indeed, using the
symmetry properties of $R_{ijkl}$  (\ref{Rs})and $T_{ijklmn}$
(\ref{Ts1}), (\ref{Ts2}), (\ref{Ts3}) we have
\begin{eqnarray}
\label{w4sR}
0&=&\omega_{ij,klmn}-\omega_{ij,lkmn}(-1)^{\epsilon_k\epsilon_l}=
\Big[R_{ikjl,mn}+R_{likj,mn}(-1)^{\epsilon_l(\epsilon_i+\epsilon_j+\epsilon_k)}+\\
\nonumber
&&R_{kjli,mn}(-1)^{\epsilon_i(\epsilon_j+\epsilon_k+\epsilon_l)}+
R_{jlik,mn}(-1)^{(\epsilon_i+\epsilon_k)(\epsilon_j+\epsilon_l)}
\Big](-1)^{\epsilon_j\epsilon_k}=\\
\nonumber
&&\Big[R_{ikjl}+R_{likj}(-1)^{\epsilon_l(\epsilon_i+\epsilon_j+\epsilon_k)}+
R_{kjli}(-1)^{\epsilon_i(\epsilon_j+\epsilon_k+\epsilon_l)}+
R_{jlik}(-1)^{(\epsilon_i+\epsilon_k)(\epsilon_j+\epsilon_l)}
\Big]_{, mn}(-1)^{\epsilon_j\epsilon_k}.
\end{eqnarray}
Due to the identity (\ref{Rjac2}) the relations (\ref{w4sR})  are
satisfied identically and there are no new identities containing
the second order affine extension of the curvature tensor.

In similar way  it is possible to find relations containing
higher order affine extensions of sypmlectic structure, the
Christoffel symbols and the curvature tensor.

\section{Higher order relations in general coordinates}

Notice, that relations (\ref{RA1}), (\ref{wAR1}), (\ref{wA3R}), (\ref{A2R}), ({\ref{Ae3l}),
(\ref{wR2})
were derived in normal coordinates. It seems to be of general
interest to find its analog relations in terms of arbitrary local
coordinates $(x)$ because the Christoffel symbols are not tensors
while the r.h.s. of (\ref{RA1}) is a tensor. It means that l.h.s.
of (\ref{RA1}) should be a tensor $G_{ijkl}$ taking the form
$\Gamma_{ijk,l}(0)$ at point $p\in M$ in normal coordinates. In
normal coordinates covariant derivative has the form of usual
partial one, but simple identification of $G_{ijkl}$  with
covariant derivative $\Gamma_{ijk;l}$ is wrong because it does not
transform accordingly tensor rules. Indeed, under change of
coordinates $(x)\rightarrow (y)$
 the Christoffel symbols $\Gamma_{ijk}$ are transformed accordingly the rule
\begin{eqnarray}
\label{Chtr0}
\nonumber
\Gamma_{ijk}(y)=
\left(\Gamma_{pqr}({x})
\frac{\partial_r {x}^r}{\partial y^k}
\frac{\partial_r {x}^q}{\partial y^j}
(-1)^{\epsilon_k(\epsilon_j+\epsilon_q)}
+ \;\omega_{pq}({x})
\frac{\partial^2_r {x}^q}{\partial y^j\partial y^k}\right)
\frac{\partial_r {x}^p}{\partial y^i}
(-1)^{(\epsilon_k+\epsilon_j)(\epsilon_i+\epsilon_p)}.
\end{eqnarray}
and, therefore, we have the following transformation law for
partial derivative of the Chistoffel symbols calculated in normal
coordinates $(y)$ via covariant derivative of the Chistoffel
symbols computed in arbitrary coordinates $(x)$
\begin{eqnarray}
\label{G1tr}
\Gamma_{ijk,l}(y)&=&\Big(\Gamma_{pqr;s}(x)-\Gamma_{pst}(x)\Gamma^{t}{}_{qr}(x)
(-1)^{\varepsilon_{s}(\varepsilon_{q}+\varepsilon_{r})}\Big)\frac{\partial_{r}x^{s}}{\partial
y^{l}}\frac{\partial_{r}x^{r}}{\partial
y^{k}}\frac{\partial_{r}x^{q}}{\partial
y^{j}}\frac{\partial_{r}x^{p}}{\partial y^{i}}\times\nonumber\\
& &
\times(-1)^{\varepsilon_{l}(\varepsilon_{i}+\varepsilon_{j}+\varepsilon_{k}
+\varepsilon_{p}+\varepsilon_{q}+\varepsilon_{r})+\varepsilon_{k}(\varepsilon_{j}+\varepsilon_{q})
+(\varepsilon_{k}+\varepsilon_{j})(\varepsilon_{i}+\varepsilon_{p})}+\nonumber\\
& &+\Gamma_{pqt}(x)\frac{\partial_{r}x^{t}}{\partial
y^{s}}\Gamma^{s}{}_{kl}(y)\frac{\partial_{r}x^{q}}{\partial
y^{j}}\frac{\partial_{r}x^{p}}{\partial
y^{i}}(-1)^{\varepsilon_{l}(\varepsilon_{i}+\varepsilon_{j}+\varepsilon_{k}+\varepsilon_{p}+\varepsilon_{q})
+\varepsilon_{k}(\varepsilon_{j}+\varepsilon_{q})+(\varepsilon_{k}+\varepsilon_{j})(\varepsilon_{i}
+\varepsilon_{p})}+\nonumber\\
& &+\Gamma_{ptr}(x)\frac{\partial_{r}x^{t}}{\partial
y^{s}}\Gamma^{s}{}_{jl}(y)\frac{\partial_{r}x^{r}}{\partial
y^{k}}\frac{\partial_{r}x^{p}}{\partial
y^{i}}(-1)^{(\varepsilon_{t}+\varepsilon_{j}+\varepsilon_{l})(\varepsilon_{r}+\varepsilon_{k})
+\varepsilon_{l}(\varepsilon_{p}+\varepsilon_{i})+\varepsilon_{k}(\varepsilon_{j}+\varepsilon_{t})+
(\varepsilon_{k}+\varepsilon_{j})(\varepsilon_{i}+\varepsilon_{p})}+\nonumber\\
& &+\Gamma_{pqs}(x)\frac{\partial_{r}x^{s}}{\partial
y^{l}}\frac{\partial_{r}x^{q}}{\partial
y^{t}}\Gamma^{t}{}_{jk}(y)\frac{\partial_{r}x^{p}}{\partial
y^{i}}(-1)^{\varepsilon_{l}(\varepsilon_{i}+\varepsilon_{j}+\varepsilon_{k}+\varepsilon_{p}+\varepsilon_{q})
(\varepsilon_{k}+\varepsilon_{j})(\varepsilon_{p}+\varepsilon_{i})}+\nonumber\\
&
&+\Gamma_{sjk}(y)\Gamma^{s}{}_{il}(y)(-1)^{(\varepsilon_{k}+\varepsilon_{j})(\varepsilon_{s}+\varepsilon_{i})}+
\nonumber\\
& &+\omega_{pq}(x)\frac{\partial_{r}^{3}x^{q}}{\partial
y^{i}\partial y^{k}\partial
y^{l}}\frac{\partial_{r}x^{p}}{\partial
y^{i}}(-1)^{\varepsilon_{l}(\varepsilon_{i}+\varepsilon_{p})+(\varepsilon_{k}+\varepsilon_{j})
(\varepsilon_{i}+\varepsilon_{p})},
\end{eqnarray}
which differs from the transformation rules of tensors on
supermanifolds (see \cite{gl2}). In (\ref{G1tr}) notation
\begin{equation}
\Gamma_{pqr;s}=\Gamma_{pqr,s}-\Gamma_{pqn}\Gamma^{n}{}_{rs}-
\Gamma_{pnr}\Gamma^{n}{}_{qs}(-1)^{\varepsilon_{k}(\varepsilon_{j}+\varepsilon_{l})}-
\Gamma_{nqr}\Gamma^{n}{}_{ps}(-1)^{(\varepsilon_{r}+\varepsilon_{q})(\varepsilon_{n}+\varepsilon_{p})},
\end{equation}
and expression for the matrix of second derivatives
\begin{equation}
\nonumber
\frac{\partial_{r}^{2}x^{q}}{\partial y^{j}\partial
y^{k}}=\frac{\partial_{r}x^{q}}{\partial
y^{l}}\Gamma^{i}{}_{jk}(y)-\Gamma^{q}{}_{lm}(x)\frac{\partial_{r}x^{m}}{\partial
y^{k}}\frac{\partial_{r}x^{l}}{\partial
y^{j}}(-1)^{\varepsilon_{k}(\varepsilon_{j}+\varepsilon_{l})}
\end{equation}
were used.

Now, making use of Eq.~(\ref{G1tr}), and
restricting to the point $p \in M$, i.e., taking  $y=0, \; x=x_0$,
we get
\begin{align}
\label{Chr1}
\Gamma_{ijk,l}(0) =
\left(\Gamma_{ijk;l}(x_0)-\Gamma_{iln}(x_0)\Gamma^n_{\;\;\;jk}(x_0)
(-1)^{\epsilon_l(\epsilon_j+\epsilon_k)}\right)+ \omega_{iq}(x_0)
\left(\frac{\partial^3_r x^q}{\partial y^j\partial y^k\partial
y^l}\right)_0.
\end{align}
Due to (\ref{Chr1}) and the identity (\ref{r3}),
the matrix of third derivatives at $p$ obeys the following
relation,
\begin{align}
\label{3w1} \nonumber \omega_{iq}(x_0) \left(\frac{\partial^3_r
x^q}{\partial y^j\partial y^k\partial y^l}\right)_0
&= \frac{1}{3}\Big[ -  \Gamma_{ijk;l} -
\Gamma_{ijl;k}(-1)^{\epsilon_k\epsilon_l} -
\Gamma_{ikl;j}(-1)^{\epsilon_j(\epsilon_k+\epsilon_l)}
\\
\nonumber
&\qquad +
\Gamma_{iln}\Gamma^n_{\;\;\;jk}(-1)^{(\epsilon_j+\epsilon_k)\epsilon_l}
+ \Gamma_{ijn}\Gamma^n_{\;\;\;kl} +
\Gamma_{ikn}\Gamma^n_{\;\;\;jl}(-1)^{\epsilon_k\epsilon_j}
\Big](x_0)\\
\qquad &=
-\frac{1}{3}\Big[Z_{ijkl}-3\Gamma_{iln}\Gamma^n_{\;\;\;jk}(-1)^{(\epsilon_k+\epsilon_j)\epsilon_l}\Big](x_0)
\end{align}
with the abbreviation
\begin{eqnarray}
\label{Zx}
Z_{ijkl}&=&
\Gamma_{ijk;l}+\Gamma_{ijl;k}(-1)^{\epsilon_k\epsilon_l}+
\Gamma_{ikl;j}(-1)^{(\epsilon_k+\epsilon_l)\epsilon_j}\\
\nonumber && +2\Gamma_{iln}\Gamma^n_{\;\;\;jk}
(-1)^{(\epsilon_k+\epsilon_j)\epsilon_l}-
\Gamma_{ikn}\Gamma^n_{\;\;\;jl}(-1)^{\epsilon_j\epsilon_k}-
\Gamma_{ijn}\Gamma^n_{\;\;\;kl}
\end{eqnarray}
With the help of (\ref{3w1}) we get the following representation for tensor $G_{ijkl}$
in arbitrary coordinates $(x)$
\begin{align}
\label{Chtr5} G_{ijkl}(x)=
 \Big[\Gamma_{ijk;l}-\frac{1}{3}Z_{ijkl}\Big](x).
\end{align}
Notice that one can express the tensor $G_{ijkl}$ in terms partial derivatives of the Christoffel symbols
as well
\begin{align}
\label{Chtr6} G_{ijkl}(x)=
 \Big[\Gamma_{ijk,l}-\frac{1}{3}X_{ijkl}\Big](x)
\end{align}
where
\begin{eqnarray}
\label{Xx}
\nonumber
X_{ijkl}&=&
\Gamma_{ijk,l}+\Gamma_{ijl,k}(-1)^{\epsilon_k\epsilon_l}+
\Gamma_{ikl,j}(-1)^{(\epsilon_k+\epsilon_l)\epsilon_j}\\
\nonumber && +2\Gamma_{njk}\Gamma^n_{\;\;\;il}
(-1)^{(\epsilon_k+\epsilon_j)(\epsilon_i+\epsilon_n)}-
\Gamma_{njl}\Gamma^n_{\;\;\;ik}(-1)^{(\epsilon_j+\epsilon_l)(\epsilon_i+\epsilon_n)+\epsilon_k\epsilon_l}\\
&&-\Gamma_{nkl}\Gamma^n_{\;\;\;ij}(-1)^{(\epsilon_k+\epsilon_l)(\epsilon_i+\epsilon_n+\epsilon_j)}.
\end{eqnarray}
Now we can find relations of the first order (\ref{RA1}), (\ref{wAR1}) in arbirtary coordinates
\begin{eqnarray}
\label{RA1x}
\Gamma_{ijk,l}-\frac{1}{3}X_{ijkl} =
-\frac{1}{3}\left[R_{ijkl}+R_{ikjl}(-1)^{\epsilon_k\epsilon_j}\right],
\end{eqnarray}
\begin{eqnarray}
\label{wAR1x}
\omega_{ij,kl}-\frac{1}{3}\Big(X_{ijkl}-X_{jikl}(-1)^{\epsilon_i\epsilon_j}\Big)
 =\frac{1}{3}R_{klij}
(-1)^{(\epsilon_i+\epsilon_j)(\epsilon_k+\epsilon_l)}.
\end{eqnarray}
The difference of $X$'s in (\ref{wAR1x}) is symmetric w.r.t two last indices because of the property
\begin{eqnarray}
\label{Xs}
\nonumber
X_{ijlk} &-&X_{ijkl}(-1)^{\epsilon_k\epsilon_l}=
3\Big[\Gamma_{njl}\Gamma^n_{\;\;\;ik}(-1)^{(\epsilon_j+\epsilon_l)(\epsilon_i+\epsilon_n)}-
\Gamma_{njk}\Gamma^n_{\;\;\;il}(-1)^{(\epsilon_k+\epsilon_j)(\epsilon_i+\epsilon_n)+\epsilon_k\epsilon_l}
\Big]=\\
&&-3\Big[\Gamma_{nik}\Gamma^n_{\;\;\;jl}(-1)^{\epsilon_n(\epsilon_i+\epsilon_k)+\epsilon_k(\epsilon_j+\epsilon_l)}+
\Gamma_{njk}\Gamma^n_{\;\;\;il}(-1)^{\epsilon_n(\epsilon_k+\epsilon_j)+
\epsilon_k(\epsilon_i+\epsilon_l)+\epsilon_i\epsilon_j} \Big].
\end{eqnarray}
It is obvious that in general coordinates the identity (\ref{IIRy}) has the form
\begin{align}
\label{IIRx} 
R_{mjik;l}(-1)^{\epsilon_j(\epsilon_i+\epsilon_k)}-
R_{mijl;k}(-1)^{\epsilon_k(\epsilon_l+\epsilon_j)}
+R_{mkjl;i}(-1)^{\epsilon_i(\epsilon_j+\epsilon_k+\epsilon_l)}-
R_{mlik;j}(-1)^{\epsilon_l(\epsilon_i+\epsilon_j+\epsilon_k)}=0.
\end{align}
It is important to note that the relation (\ref{wA3R}) can be derived from (\ref{wAR1x})
written in normal coordinates by differentiation w.r.t. $y$ and then putting $y=0$.

By differentiation of (\ref{G1tr}) w.r.t. $y$ we obtain the
expression for the second order affine extension of the
Christoffel symbols
\begin{eqnarray}
\label{G2}
\Gamma_{ijk,lm}(0)&=&\Gamma_{ijk;lm}(x_{0})-\Gamma_{ils;m}(x_{0})\Gamma^{s}{}_{jk}(x_{0})
(-1)^{\varepsilon_{m}(\varepsilon_{s}+\varepsilon_{j}+\varepsilon_{k})+
\varepsilon_{l}(\varepsilon_{j}+\varepsilon_{k})}+\nonumber\\
& &
+\Gamma_{ijs}(x_{0})\Gamma^{s}{}_{kl,m}(0)+\Gamma_{isk}(x_{0})\Gamma^{s}{}_{jl,m}(0)
(-1)^{\varepsilon_{k}(\varepsilon_{j}+\varepsilon_{s})}-\nonumber\\
&&
-\frac{1}{3}\Gamma_{isl}(x_{0})Z^{s}{}_{jkm}(x_{0})(-1)^{\varepsilon_{l}(\varepsilon_{j}+
\varepsilon_{k}+\varepsilon_{s})}+\nonumber\\
& &
+\Gamma_{ism}(x_{0})\left(\frac{\partial_{r}^{3}x^{s}}{\partial
y^{j}\partial y^{k}\partial
y^{l}}\right)_{0}+\omega_{is}(x_{0})\left(\frac{\partial_{r}^{4}x^{s}}{\partial
y^{j}\partial y^{k}\partial y^{l}\partial y^{m}}\right)_{0}.
\end{eqnarray}
Using (\ref{G2}) and identity (\ref{r6}) one can find a closed
form of the matrix of forth derivatives at $p$ and therefore
relations  (\ref{Ae3l}) and (\ref{wR2}) in general coordinates.
Here we do not give explicit formulas for them restricting ourself
by explicit expressions for the matrix of third derivatives and
for $\omega_{ij,kl}$ only.

\section{Discussion}

We have considered  properties of Fedosov supermanifolds. Using
normal coordinates on a supermanifold we have found relations up
to the third order among the  affine extensions of the Christoffel
symbols and the curvature tensor,  the  affine extensions of
symplectic structure and the curvature tensor as well as
identities for the curvature tensor. Considering relations of the
third order it was checked absence of independent identities
containing the second order (covariant) derivatives of the
curvature tensor. It was shown principal role of tensor
$G_{ijkl}=\Gamma_{ijk,l}-1/3X_{ijkl}$ to obtain the relations in
general local coordinates. In fact the relations (\ref{wAR1x})
should be considered as the fundamental ones to derive  step by
step all higher order relations by covariant differentiation of
its right and left sides. Notice that the tensor of such kind is
not specific for symplectic geometry only and can be introduced in
both affine and Riemannian geometries too. Indeed, let us
introduced the quantity
\begin{eqnarray}
\label{Gg}
G^i_{\;\;jkl}=\Gamma^i_{\;\;jk;l}-\frac{1}{3}Z^i_{\;\;jkl}=
\Gamma^i_{\;\;jk,l}-\frac{1}{3}Y^i_{\;\;jkl}
\end{eqnarray}
where $\Gamma^i_{\;\;jk}$ is an affine connection on a
supermanifold and notations
\begin{eqnarray}
\label{ZXg} \nonumber
Z^i_{\;\;jkl}&=&
\Gamma^i_{\;\;jk;l}+\Gamma^i_{\;\;jl;k}(-1)^{\epsilon_k\epsilon_l}+
\Gamma^i_{\;\;kl;j}(-1)^{(\epsilon_k+\epsilon_l)\epsilon_j}\\
&& +2\Gamma^i_{\;\;ln}\Gamma^n_{\;\;\;jk}
(-1)^{(\epsilon_k+\epsilon_j)\epsilon_l}-
\Gamma^i_{\;\;kn}\Gamma^n_{\;\;\;jl}(-1)^{\epsilon_j\epsilon_k}-
\Gamma^i_{\;\;jn}\Gamma^n_{\;\;\;kl},\\
\nonumber
Y^i_{\;\;jkl}&=&
\Gamma^i_{\;\;jk,l}+\Gamma^i_{\;\;jl,k}(-1)^{\epsilon_k\epsilon_l}+
\Gamma^i_{\;\;kl,j}(-1)^{(\epsilon_k+\epsilon_l)\epsilon_j}
-2\Gamma^i_{\;\;ln}\Gamma^n_{\;\;\;jk}
(-1)^{\epsilon_l(\epsilon_j+\epsilon_k)}+\\
&&\Gamma^i_{\;\;kn}\Gamma^n_{\;\;\;jl}(-1)^{\epsilon_k\epsilon_j}+
\Gamma^i_{\;\;jn}\Gamma^n_{\;\;\;kl},\\
\Gamma^i_{\;\;jk;l}&=&\Gamma^i_{\;\;jk,l}-
\Gamma^i_{\;\;jn}\Gamma^n_{\;\;\;kl}
-\Gamma^i_{\;\;kn}\Gamma^n_{\;\;\;jl} (-1)^{\epsilon_k\epsilon_j}+
\Gamma^i_{\;\;ln}\Gamma^n_{\;\;\;jk}(-1)^{\epsilon_l(\epsilon_j+\epsilon_k)}
\end{eqnarray}
are used. Then we get representation of the curvature tensor in
terms of $G^i_{\;\;jkl}$
\begin{eqnarray}
\label{Raf}
R^i_{\;\;jkl}=-G^i_{\;\;jkl}+G^i_{\;\;jlk}(-1)^{\epsilon_l\epsilon_k},
\end{eqnarray}
and of $G^i_{\;\;jkl}$ in terms of the curvature tensor
\begin{eqnarray}
\label{Gaf}
G^i_{\;\;jkl}=-\frac{1}{3}\big[R^i_{\;\;jkl}+R^i_{\;\;kjl}(-1)^{\epsilon_j\epsilon_k}\big],
\end{eqnarray}
which proves tensor character of $G^i_{\;\;jkl}$.
In the Riemannian geometry a metric tensor $g_{ij}$ on supermanifold $M$ is additionally introduced.
This tensor is symmetric
\begin{eqnarray}
\label{g}
g_{ij}=g_{ji}(-1)^{\epsilon_i\epsilon_j},
\end{eqnarray}
and covariant constant
\begin{eqnarray}
\label{gcon}
g_{ij,k}=g_{im}\Gamma^m_{\;\;jk}+g_{jm}\Gamma^m_{\;\;ik}(-1)^{\epsilon_i\epsilon_j}.
\end{eqnarray}
With the help of $g_{ij}$ one can lower the upper index of $R^i_{\;\;jkl}$ to obtain the
tensor $R_{ijkl}$
\begin{eqnarray}
\nonumber
\label{Rlg}
R_{ijkl}=g_{im}R^m_{\;\;jkl},
\end{eqnarray}
obeying the following symmetry properties
\begin{eqnarray}
\label{Rsym}
R_{ijkl}=-R_{jikl}(-1)^{\epsilon_i\epsilon_j}, \quad
R_{ijkl}=R_{klij}(-1)^{(\epsilon_i+\epsilon_j)(\epsilon_k+\epsilon_l)}.
\end{eqnarray}
In turn we have
\begin{eqnarray}
\label{Gl}
G_{ijkl}=g_{im}G^m_{\;\;jkl}=\Gamma_{ijk;l}-\frac{1}{3}Z_{ijkl}=\Gamma_{ijk,l}-\frac{1}{3}X_{ijkl},
\end{eqnarray}
where $\Gamma_{ijk}=g_{im}\Gamma^m_{\;\;jk},\;\;
Z_{ijkl}=g_{im}Z^m_{\;\;jkl}$  and the expression for $X_{ijkl}$
formally coinsides with (\ref{Xx}). From (\ref{Raf}), (\ref{Gaf}),
(\ref{gcon}) it follows
\begin{eqnarray}
\label{Rrim}
R_{ijkl}=-G_{ijkl}+G_{ijlk}(-1)^{\epsilon_l\epsilon_k},
\quad
G_{ijkl}=-\frac{1}{3}\big[R_{ijkl}+R_{ikjl}(-1)^{\epsilon_j\epsilon_k}\big],
\quad
g_{ij,k}=\Gamma_{ijk}+\Gamma_{jik}(-1)^{\epsilon_i\epsilon_j}.
\end{eqnarray}
In similar manner used in Section 3  we can introduce normal
coordinates on a supermanifold $M$ and derive the following
relation
\begin{eqnarray}
\label{gsec}
g_{ij,kl}(0)=-\frac{1}{3}\big[R_{ikjl}(0)(-1)^{\epsilon_j(\epsilon_i+\epsilon_k)}+R_{jkil}(0)
(-1)^{\epsilon_i\epsilon_k}\big],
\end{eqnarray}
which can be considered as an analog of (\ref{wAR1}) in the
Riemannian geometry. It is easily to check that due to
(\ref{Rsym}) symmetry properties of  $g_{ij,kl}(0)$ are in
accordance with this relation. In general coordinates this
relation has the form
\begin{eqnarray}
\label{gsecx}
g_{ij,kl}-\frac{1}{3}\big[X_{ijkl}+X_{jikl}(-1)^{\epsilon_i\epsilon_j}\big]
=-\frac{1}{3}\big[R_{ikjl}(-1)^{\epsilon_j(\epsilon_i+\epsilon_k)}+R_{jkil}
(-1)^{\epsilon_i\epsilon_k}\big].
\end{eqnarray}
From (\ref{gsecx}) higher order relations can be obtained by
covariant differentiations.

\medskip
\vspace{0.5cm}

\noindent
{\Large{\bf {Acknowledgments}}}

\vspace{0.5cm} \noindent We would like to thank I.L. Buchbinder,
B. Geyer, A.P. Nersessian, V.V. Obukhov, K.E. Osetrin, D.V.
Vassilevich and I.V. Volovich  for useful discussions. The work of PLM was supported
by Deutsche Forschungsgemeinschaft (DFG) grant DFG 436 RUS
113/669/0-2, Russian Foundation for Basic Research (RFBR) grants
03-02-16193 and 04-02-04002, the President grant LSS 1252.2003.2
and INTAS grant 03-51-6346.

\end{document}